\documentstyle[amsfonts,aps]{revtex}

\begin{document}
\title{Subdynamics theory in the functional approach to quantum mechanics}
\author{Roberto Laura and Rodolfo M. Id Betan}
\address{Departamenteo de F\'{\i}sica, F.C.E.I.A. Universidad Nacional de Rosario.\\
Instituto de F\'{\i}sica Rosario, CONICET-UNR.\\
Av. Pellegrini 250, 2000 Rosario, Argentina\\
e-mail: laura@ifir.ifir.edu.ar}
\maketitle

\begin{abstract}
The formalism of subdynamics is extended to the functional approach of
quantum systems, and used for the Friedrichs model, in which diagonal
singularities in states and observables are included. We compute in this
approach the generalized eigenvectors and eigenvalues of the Liouvulle-Von
Newmann operator, using an iterative scheme. As complex generalized
eigenvalues are obtained, the decay rates of unstable modes are included in
the spectral decomposition.
\end{abstract}

\section{Introduction}

For quantum system with continuous spectrum, the presence of resonances
(small denominators) cause the failure of the usual perturbative methods for
computing eigenvalues and eigenvectors of the time evolution generator.
These difficulties have been considered as a manifestation of general
limitations to computability for unstable dynamical systems\cite{1}\cite{2}.

In the work of the Brussels-Austin groups on Large Poincare systems, we find
an algorithm to overcome the problem of small denominators, which are
eliminated by a ''time ordering rule''. This is a rule for the
regularization of the perturbation terms, which can be interpreted as a
generalized boundary condition where terms corresponding to excitation
processes are past-oriented, while terms corresponding to the de-excitation
and emission of radiation are future-oriented\cite{1}\cite{3}.

The construction provides a new type of spectral decomposition of the
Hamiltonian operator. For the Friedrichs model it was shown \cite{1} that an
appropriate mathematical framework for the time ordering construction is the
theory of rigged Hilbert spaces of the Hardy class, formulated by A.Bohm and
M.Gadella\cite{4}\cite{5}\cite{6}\cite{7}.

For mixed states, the construction algorithm is a generalization of
perturbation theory based on the subdynamics decomposition of the
Liouville-Von Newmann superoperator ${\Bbb L}$ \cite{2}\cite{8}. Through a
non unitary transformation $\Omega ,$ the superoperator ${\Bbb L}$ is
transformed into an operator $\Theta =\Omega ^{-1}{\Bbb L}\Omega $, which is
block diagonal in the degrees of correlations. The perturbative method to
obtain the spectral decomposition of the intermediate operator $\Theta $ is
regularized by imposing the ''time-ordering rules'', which in this case
means that conserving or increasing of degrees of correlations is future
oriented, while decreasing of correlations is past oriented\cite{9}. This
prescription explicitly incorporates irreversibility to microscopic theories.

Usually, these perturbative algoritms are implemented in the so called ''box
normalization'', in which the quantum system is assumed to be included in a
box with periodic boundary conditions, the size of the box becoming infinite
at some stage of the calculations. To perform this limit, it is necessary to
consider volume dependent factors both for the diagonal components of the
density operator and the observables. In this limit the recurrence time of
the system is pushed to infinity.

The diagonal singularities of operators for large quantum system was
discovered by Van Howe \cite{10}\cite{11}\cite{12}\cite{13}. At the same
time, I. Prigogine and coworkers \cite{14}\cite{15}\cite{16}\cite{17}
emphasized the importance of states with diagonal singularity in non
equilibrium statistical physics.

Based in the pioneering work of I. E. Segal \cite{18}, I. Antoniou et al 
\cite{19}\cite{20}\cite{21} developed a formalism for quantum systems with
continuous spectrum without the box normalization.

The quantum states are {\it functionals} over certain space of observables $%
{\cal O}$. Mathematically this means that the space ${\cal S}$ of states is
contained in ${\cal O}^{\times }.$ Physically it means that the only thing
we can really observe and measure are the mean values of the observables $%
O\in {\cal O}$ in states $\rho \in {\cal S\subset }{\cal O}^{\times }$ ($%
\langle O\rangle _\rho =(\rho \mid O)$). This is the natural generalization
of the usual trace of the product of the density operator by the observable (%
$Tr(\widehat{\rho }\widehat{O})$) which is not well defined for systems with 
{\it continuous spectrum.}

In this paper, we extend the theory of subdynamics to the case of quantum
systems with diagonal singularities where, as stated in references \cite{19}%
\cite{20}\cite{21}, the states are considered as functionals acting on the
space of observables. The extended formalism is applied to the Friedrichs
Model.

In section II, we summarize the functional approach to quantum mechanics.
The theory of subdynamics \cite{2}\cite{8}\cite{9} is summarized in section
III, and it is extended by us to the functional approach. In section IV, the
extended formalism is applied to compute the generalized spectral
decomposition and the time evolution for the Friedrichs Model.

\section{Functional Approach To Quantum Mechanics}

In the usual approach to quantum mechanics, a pure state is represented by a
normalized vector $|\psi \rangle $ of a Hilbert space ${\cal H}$. The
observables of the system are represented by self adjoint operators acting
on ${\cal H}$. The mean value of an observable represented by the operator $%
O $ in a pure state represented by the vector $|\Psi \rangle $ is given by 
\[
\langle O\rangle _\Psi =\langle \Psi |O|\Psi \rangle 
\]

The time evolution of the state vector is given by the Schr\"{o}dinger
equation 
\[
i\frac d{dt}|\Psi _t\rangle =H\,|\Psi _t\rangle , 
\]
where the operator $H:{\cal H}{\em \longrightarrow }{\cal H}$ is the
Hamiltonian operator of the system. Schr\"{o}dinger equation has the
solution 
\[
|\Psi _t\rangle =e^{-iHt}\,|\Psi _0\rangle 
\]

Mixed states have no well defined state vectors, but a probabilities $%
p_\alpha $ ($p_\alpha \geq 0$, $\sum_\alpha p_\alpha =1$) of being in the
pure states represented by normalized vectors $|\Psi _\alpha \rangle $.
Therefore, the mean value of an observable $O$ is given by 
\[
\left\langle O\right\rangle =\sum\limits_\alpha p_\alpha \langle \Psi
_\alpha |O|\Psi _\alpha \rangle . 
\]

The mixed state can be represented by the {\it density operator} $\widehat{%
\rho }=\sum_\alpha p_\alpha |\Psi _\alpha \rangle \langle \Psi _\alpha |,$
having the following properties 
\[
\langle O\rangle =Tr(\widehat{\rho }O),\qquad Tr(\widehat{\rho }%
)=\sum\limits_\alpha p_\alpha \langle \Psi _\alpha |\Psi _\alpha \rangle
=\sum\limits_\alpha p_\alpha =1 
\]

As each vector $|\Psi _\alpha \rangle $ evolves in time according to the
Schr\"{o}dinger equation, the time evolution of $\widehat{\rho }$ is 
\[
\widehat{\rho }_t=e^{-iHt}\,\widehat{\rho }_0\,e^{iHt} 
\]
and $\widehat{\rho }_t$ satisfies the Liouville-Von Newmann equation 
\begin{equation}
i\frac d{dt}\widehat{\rho }_t=L\,\widehat{\rho }_t  \label{2-1}
\end{equation}
\begin{equation}
L\widehat{\rho }\equiv [H,\widehat{\rho }]  \label{2-2}
\end{equation}

In a more general approach, {\it the set of all possible observables of a
quantum system is represented by an algebra }${\cal O}${\it , while the
possible states are represented by a set }${\cal S}${\it \ of functionals
acting on }${\cal O}${\it \ (}${\cal S}\subset {\cal O}^{\times }${\it ).}

The mean value $\langle O\rangle _\rho $ of the observable $O$ in the state $%
\rho $ is given by the value of the functional $\rho $ on $O$, which we
denote by $(\rho |O)$%
\[
\langle O\rangle _\rho =(\rho |O) 
\]

The last expression is antilinear in $\rho $ and linear in $O,$ i.e. 
\begin{equation}
(\alpha _1\rho _1+\alpha _2\rho _2|O)=\alpha _1^{*}(\rho _1|O)+\alpha
_2^{*}(\rho _2|O)  \label{2-3}
\end{equation}
\begin{equation}
(\rho |\alpha _1O_1+\alpha _2O_2)=\alpha _1(\rho |O_1)+\alpha _2(\rho |O_2)
\end{equation}

The algebra ${\cal O}$ is chosen to be an algebra of self adjoint operators
on the vector space ${\cal H}$, and as the mean value of the observables
should be a real number, we impose the following condition on the states 
\begin{equation}
(\rho |O)=(\rho |O)^{*},\text{ if }O=O^{\dagger }
\end{equation}

The {\it generalization of the concept of trace} is 
\begin{equation}
Tr\,\rho \equiv (\rho |I)=1
\end{equation}
where $I$ is the identity operator in the algebra ${\cal O}$.

For the time evolution in Heisenberg representation, the states are time
independent, while the observables evolve in time according to 
\[
O_t=e^{i\,H\,t}Oe^{-i\,H\,t}. 
\]

The time evolution $\rho _t$ of the states in Schr\"{o}dinger representation
can be obtained from 
\[
\langle O\rangle _t=(\rho _t|O)=(\rho _0|e^{i\,H\,t}Oe^{-i\,H\,t}). 
\]

From the previous equation we obtain 
\[
(\frac d{dt}\rho _t|O)=i(\rho _0|e^{i\,H\,t}\,[H,O]\,e^{-i\,H\,t})=i(\rho
_t|[H,O]) 
\]

Calling ${\Bbb L}^{\dagger }O\equiv [H,O]$ and omitting the observable $O$
in the previous equation we obtain the generalized Liouville-Von Newmann
equation 
\begin{equation}
-i(\frac d{dt}\rho _t|=(\rho _t|{\Bbb L}^{\dagger }  \label{2-7}
\end{equation}

To each superoperator ${\Bbb M}$ acting on ${\cal S}$ we can associate a
corresponding adjoint superoperator ${\Bbb M}^{\dagger }$ acting on ${\cal O}
$ and viceversa, with the formula 
\begin{equation}
({\Bbb M}\rho |O)=(\rho |{\Bbb M}^{\dagger }O).  \label{2-8}
\end{equation}

Using equation (\ref{2-8}) with ${\Bbb M}={\Bbb L},$ and the antilinearity
property (\ref{2-3}), in equation (\ref{2-7}), we obtain 
\[
i\frac d{dt}\rho _t={\Bbb L\,}\rho _t, 
\]
which is formally the same as equation (\ref{2-1}). However, as $\rho $ {\it %
is not an operator}, equation (\ref{2-2}) is not more valid. The extended
Liouville-Von Newmann operator ${\Bbb L}$ acting on ${\cal S}$ is now given
by 
\begin{equation}
({\Bbb L\,}\rho |O)=(\rho |{\Bbb L}^{\dagger }O)\equiv (\rho |[H,O]).
\end{equation}

\section{The formalism of Subdynamics}

Let us consider a linear space of states ${\cal S}$, and a linear operator $%
{\Bbb L}$ on ${\cal S}$ which is the generator of the time evolution of the
states, i.e. 
\begin{equation}
i\frac d{dt}\rho _t={\Bbb L\,}\rho _t,\,\,\;\;\;\;\;\;\rho _t\in {\cal S}.
\label{3-1}
\end{equation}

Let us assume that the operator ${\Bbb L}$ can be decomposed into 
\begin{equation}
{\Bbb L}={\Bbb L}_0+{\Bbb L}_1
\end{equation}
where ${\Bbb L}_0$ and ${\Bbb L}_1$ are respectively called the ''free'' and
''interaction'' parts of ${\Bbb L}$. It is also assumed that an interaction
parameter $\lambda $ is included in ${\Bbb L}_1$ to modulate the interaction.

Starting with the projector ${\Bbb P}_0$ on the invariant parts of the
dynamics (${\Bbb L}_0{\Bbb P}_0={\Bbb P}_0{\Bbb L}_0=0$), projectors ${\Bbb P%
}_n$ ($n=0,1,...$) are defined in such a way that they satisfy 
\begin{equation}
{\Bbb P}_n{\Bbb P}_{n^{^{\prime }}}=\delta _{nn^{^{\prime }}}{\Bbb P}%
_n,\quad {\Bbb L}_0{\Bbb P}_n={\Bbb P}_n{\Bbb L}_0,\quad \sum_{n=0}{\Bbb P}%
_n={\Bbb I},\quad {\Bbb P}_m\left( {\Bbb L}_1\right) ^n{\Bbb P}_0=\left\{ 
\begin{tabular}{ll}
$=0$ & $if\;n<m$ \\ 
$\neq 0$ & $if\;n=m$%
\end{tabular}
\right.  \label{3-3}
\end{equation}
where ${\Bbb I}$ is the identity operator on ${\cal S}$. The last equation
means that the transition from ${\Bbb P}_0\rho $ to ${\Bbb P}_m\rho $ is a
process of m-th order in the interaction parameter. The operator ${\Bbb P}_m$
es called the {\it projection on the m-th degree of correlation.}

The main idea of subdynamics is to decompose the states through projectors $%
\Pi _n$ ($n=0,1,...)$ satisfying

\begin{equation}
\Pi _n\Pi _{n^{^{\prime }}}=\delta _{nn^{^{\prime }}}\Pi _n,\quad {\Bbb L}%
\Pi _n=\Pi _n{\Bbb L},\quad \sum_{n=0}\Pi _n={\Bbb I},\quad \lim_{{\Bbb L}%
_1\rightarrow 0}\Pi _n={\Bbb P}_n,
\end{equation}
i.e. projectors $\Pi _n$ commuting with ${\Bbb L}$ which reduce to the
projectors ${\Bbb P}_n$ on the degrees of correlation when the parameter of
the interaction tends to zero.

Operators ${\Bbb C}_n$ and ${\Bbb D}_n$, called creation and destruction of
correlations are defined by 
\begin{eqnarray}
{\Bbb C}_n &=&{\Bbb Q}_n{\Bbb C}_n{\Bbb P}_n\;\;\;\;\;\;{\Bbb Q}_n\Pi _n=%
{\Bbb C}_n{\Bbb P}_n\Pi _n  \nonumber \\
{\Bbb D}_n &=&{\Bbb P}_n{\Bbb D}_n{\Bbb Q}_n\;\;\;\;\;\;\Pi _n{\Bbb Q}_n=\Pi
_n{\Bbb P}_n{\Bbb D}_n  \label{3-5}
\end{eqnarray}

where ${\Bbb Q}_n={\Bbb I}-{\Bbb P}_n.$ From equations (\ref{3-1}) and (\ref
{3-5}) it is obtained

\begin{eqnarray}
i\frac d{dt}({\Bbb P}_n\Pi _n\rho ) &=&\Theta _n{\Bbb P}_n\Pi _n\rho \\
\Theta _n &=&{\Bbb P}_n{\Bbb LP}_n+{\Bbb P}_n{\Bbb LC}_n{\Bbb P}_n
\label{3-7} \\
\Pi _n &=&({\Bbb P}_n+{\Bbb C}_n)({\Bbb P}_n+{\Bbb D}_n{\Bbb C}_n)^{-1}(%
{\Bbb P}_n+{\Bbb D}_n) \\
\lbrack {\Bbb L}_0,{\Bbb P}_m{\Bbb C}_n] &=&({\Bbb P}_m{\Bbb C}_n-{\Bbb P}_m)%
{\Bbb L}_1({\Bbb P}_n+{\Bbb C}_n) \\
\lbrack {\Bbb L}_0,{\Bbb D}_n{\Bbb P}_m] &=&({\Bbb P}_n+{\Bbb D}_n){\Bbb L}%
_1({\Bbb P}_m-{\Bbb D}_n{\Bbb P}_m).
\end{eqnarray}

The last two equations have the form $[{\Bbb L}_0,{\Bbb X}]={\Bbb Y}$,
having the forward (backward) solutions 
\[
{\Bbb X}^{\pm }=i\int\limits_0^{\pm \infty }dt\;e^{-i{\Bbb L}_0\,t}\,{\Bbb Y}%
\,e^{i{\Bbb L}_0t} 
\]

The following ''time ordering rule'' is chosen: +(-) sign is used for ${\Bbb %
X}={\Bbb P}_m{\Bbb C}_n$ with $m>n$ ($m<n$), and for ${\Bbb X}={\Bbb D}_n%
{\Bbb P}_m$ with $n>m$ ($n<m$). Therefore, the creation (destruction) of
correlations, is future (past) oriented, i.e. 
\begin{eqnarray}
{\Bbb P}_m{\Bbb C}_n &=&i\int\limits_0^{\pm \infty }dt\,e^{-i{\Bbb L}_0t}\,(%
{\Bbb P}_m{\Bbb C}_n-{\Bbb P}_m)\,{\Bbb L}_1\,({\Bbb P}_n+{\Bbb C}_n)\,e^{i%
{\Bbb L}_0t},\qquad m%
{> \atop <}
n  \nonumber \\
{\Bbb D}_n{\Bbb P}_m &=&i\int\limits_0^{\pm \infty }dt\,e^{-i{\Bbb L}_0t}\,(%
{\Bbb P}_n+{\Bbb D}_n)\,{\Bbb L}_1\,({\Bbb P}_m-{\Bbb D}_n{\Bbb P}_m)\,e^{i%
{\Bbb L}_0t}\qquad m%
{\textstyle {< \atop >}}
n  \label{3-11}
\end{eqnarray}

The two previous equations can be solved iteratively to the required order
in the interaction parameter, starting with the zero order solutions 
\[
{\Bbb C}_n^{\left( 0\right) }={\Bbb D}_n^{\left( 0\right) }=0. 
\]

Once ${\Bbb C}_n$ and ${\Bbb D}_n$ are obtained, the intermediate
superoperators $\Theta _n={\Bbb P}_n\Theta _n{\Bbb P}_n$ can be computed
using (\ref{3-7}). The block diagonal super operator $\Theta
=\sum\limits_n\Theta _n$ satisfies 
\begin{eqnarray}
{\Bbb L} &=&\Omega \,\Theta \,\Omega ^{-1},  \nonumber \\
\Omega &=&\sum\limits_n\left( {\Bbb P}_n+{\Bbb C}_n\right) ,  \label{3-12} \\
\Omega ^{-1} &=&\sum\limits_n\left( {\Bbb P}_n+{\Bbb D}_n{\Bbb C}_n\right)
^{-1}\left( {\Bbb P}_n+{\Bbb D}_n\right) ,  \nonumber
\end{eqnarray}
and therefore it is isospectral with the Liouville-Von Newmann operator $%
{\Bbb L}$. This property can be used to obtain the spectral decomposition of 
${\Bbb L}$ in terms of the spectral decomposition of the superoperator $%
\Theta $, with the same generalized eigenvalues.

The formalism of subdynamics originally stated on the space of density
operators can be easily translated to the functional approach of quantum
mechanics described in the previous section. All the formulas of this
section are still valid, but we should remember that the superoperators are
in this case defined on the space ${\cal S}$ of functionals.

It is operationally more convenient to rewrite all the previous equations
for the corresponding adjoint superoperators, acting on the space of
observables ${\cal O}.$ This is easily done by using the adjoint relation 
\[
(\alpha \,{\Bbb M\,N\,\rho }|O)=({\Bbb \rho }|\alpha ^{*}{\Bbb N}^{\dagger }%
{\Bbb M}^{{\Bbb \dagger }}O). 
\]

For example, equations (\ref{3-3}) for the projections on the degrees of
correlations are replaced by 
\begin{equation}
{\Bbb P}_n^{\dagger }{\Bbb P}_{n^{^{\prime }}}^{\dagger }=\delta
_{nn^{^{\prime }}}{\Bbb P}_n^{\dagger },\,\;\;\;{\Bbb L}_0^{\dagger }{\Bbb P}%
_n^{\dagger }={\Bbb P}_n^{\dagger }{\Bbb L}_0^{\dagger
},\;\;\;\;\sum\limits_n{\Bbb P}_n^{\dagger }={\Bbb I}^{{\Bbb \dagger }%
},\,\;\;\;\;{\Bbb P}_0^{\dagger }({\Bbb L}_1^{\dagger })^n{\Bbb P}%
_m^{\dagger }=\left\{ 
\begin{tabular}{ll}
$=0$ & $if\;n<m$ \\ 
$\neq 0$ & $if\;n=m$%
\end{tabular}
\right. ,
\end{equation}
where ${\Bbb I}^{\dagger }$ is the identify superoperator on ${\cal O}.$

Equations (\ref{3-11}) for the creation and destruction of correlations
transform into 
\begin{eqnarray}
{\Bbb C}_n^{{\Bbb \dagger }}{\Bbb P}_m^{{\Bbb \dagger }}
&=&-i\int\limits_0^{\pm \infty }dt\,e^{-i{\Bbb L}_0^{{\Bbb \dagger }}t}\,(%
{\Bbb P}_n^{{\Bbb \dagger }}+{\Bbb C}_n^{{\Bbb \dagger }})\,{\Bbb L}_1^{%
{\Bbb \dagger }}\,({\Bbb C}_n^{{\Bbb \dagger }}{\Bbb P}_m^{{\Bbb \dagger }}-%
{\Bbb P}_m^{{\Bbb \dagger }})\,e^{i{\Bbb L}_0^{{\Bbb \dagger }}t},\qquad m%
{> \atop <}
n  \label{3-14} \\
{\Bbb P}_m^{{\Bbb \dagger }}{\Bbb D}_n^{{\Bbb \dagger }}
&=&-i\int\limits_0^{\pm \infty }dt\,e^{-i{\Bbb L}_0^{{\Bbb \dagger }}t}\,(%
{\Bbb P}_m^{{\Bbb \dagger }}-{\Bbb P}_m^{{\Bbb \dagger }}{\Bbb D}_n^{{\Bbb %
\dagger }})\,{\Bbb L}_1^{{\Bbb \dagger }}\,({\Bbb P}_n^{{\Bbb \dagger }}+%
{\Bbb D}_n^{{\Bbb \dagger }})\,e^{i{\Bbb L}_0^{{\Bbb \dagger }}t},\qquad m%
{\textstyle {< \atop >}}
n,  \label{3-15}
\end{eqnarray}
where now 
\begin{equation}
{\Bbb C}_n^{{\Bbb \dagger }}={\Bbb P}_n^{{\Bbb \dagger }}{\Bbb C}_n^{{\Bbb %
\dagger }}{\Bbb Q}_n^{{\Bbb \dagger }},\;\;\;\;\;{\Bbb D}_n^{{\Bbb \dagger }%
}={\Bbb Q}_n^{{\Bbb \dagger }}{\Bbb D}_n^{{\Bbb \dagger }}{\Bbb P}_n^{{\Bbb %
\dagger }},\;\;\;\;\;{\Bbb Q}_n^{{\Bbb \dagger }}={\Bbb I}^{{\Bbb \dagger }}-%
{\Bbb P}_n^{{\Bbb \dagger }}
\end{equation}

Equations (\ref{3-14}) an (\ref{3-15}) can be written in a form which is
more suitable for calculations. Let us consider the generalized left and
right eigenvectors $(\alpha |$ and $|\beta )$ of ${\Bbb L}_0^{\dagger },$
having degrees of correlation $n_\alpha $ and $n_\beta ,$ i.e. 
\[
(\alpha |{\Bbb L}_0^{\dagger }=\omega _\alpha (\alpha |,\qquad {\Bbb L}%
_0^{\dagger }|\beta )=\omega _\beta |\beta ),\qquad (\alpha |=(\alpha |{\Bbb %
P}_{n_\alpha }^{\dagger },\qquad |\beta )={\Bbb P}_{n_{{}_\beta }}^{\dagger
}|\beta ). 
\]

From equation (\ref{3-14}) we obtain 
\[
(\alpha |{\Bbb C}_n^{{\Bbb \dagger }}|\beta )=-i\int\limits_0^{\pm \infty
}dt\,e^{i(\omega _\beta -\omega _\alpha )\,t}\,(\alpha |({\Bbb P}_n^{{\Bbb %
\dagger }}+{\Bbb C}_n^{{\Bbb \dagger }})\,{\Bbb L}_1^{{\Bbb \dagger }}\,(%
{\Bbb C}_n^{{\Bbb \dagger }}-{\Bbb Q}_n^{{\Bbb \dagger }})\,|\beta ),\qquad
n_\beta 
{> \atop <}
n_\alpha 
\]

If we use in the previous expression the identity 
\[
\int\limits_0^{\pm \infty }dt\,\,e^{ixt}=\frac i{x\pm i0} 
\]

we obtain 
\begin{equation}
(\alpha |{\Bbb C}_n^{{\Bbb \dagger }}|\beta )=\frac 1{\omega _\beta -\omega
_\alpha \pm i0}(\alpha |\,({\Bbb P}_n^{{\Bbb \dagger }}+{\Bbb C}_n^{{\Bbb %
\dagger }})\,{\Bbb L}_1^{{\Bbb \dagger }}\,({\Bbb C}_n^{{\Bbb \dagger }}-%
{\Bbb Q}_n^{{\Bbb \dagger }})\,|\beta ),\qquad n_\beta 
{> \atop <}
n_\alpha  \label{3-20}
\end{equation}

In the same way we obtain 
\begin{equation}
(\alpha |{\Bbb D}_n^{{\Bbb \dagger }}|\beta )=\frac 1{\omega _\beta -\omega
_\alpha \pm i0}(\alpha |\,({\Bbb Q}_n^{{\Bbb \dagger }}-{\Bbb D}_n^{{\Bbb %
\dagger }})\,{\Bbb L}_1^{{\Bbb \dagger }}\,({\Bbb P}_n^{{\Bbb \dagger }}+%
{\Bbb D}_n^{{\Bbb \dagger }})\,|\beta ),\qquad n_\beta 
{< \atop >}
n_\alpha  \label{3-21}
\end{equation}

The intermediate operator $\Theta ^{\dagger }$ is 
\begin{equation}
\Theta ^{\dagger }=\sum\limits_n\Theta _n^{\dagger },\;\;\;\;\;\;\;\Theta
_n^{\dagger }={\Bbb P}_n^{\dagger }{\Bbb L}^{\dagger }{\Bbb P}_n^{\dagger }+%
{\Bbb P}_n^{\dagger }{\Bbb C}_n^{\dagger }{\Bbb L}^{\dagger }{\Bbb P}%
_n^{\dagger }  \label{3-22}
\end{equation}

The intermediate operator $\Theta ^{\dagger }$ is isospectral with ${\Bbb L}%
^{\dagger }$%
\begin{eqnarray}
{\Bbb L}^{\dagger } &=&(\Omega ^{\dagger })^{-1}\Theta ^{\dagger }\Omega
^{\dagger }  \nonumber \\
\Omega ^{\dagger } &=&\sum\limits_n({\Bbb P}_n^{\dagger }+{\Bbb C}%
_n^{\dagger })  \label{3-20a} \\
(\Omega ^{\dagger })^{-1} &=&\sum\limits_n({\Bbb P}_n^{\dagger }+{\Bbb D}%
_n^{\dagger })({\Bbb P}_n^{\dagger }+{\Bbb C}_n^{\dagger }{\Bbb D}%
_n^{\dagger })^{-1}  \nonumber
\end{eqnarray}

For $({\Bbb P}_n^{\dagger }+{\Bbb C}_n^{\dagger }{\Bbb D}_n^{\dagger })^{-1}$
we can use the following expansion 
\begin{equation}
({\Bbb P}_n^{\dagger }+{\Bbb C}_n^{\dagger }{\Bbb D}_n^{\dagger })^{-1}=%
{\Bbb P}_n^{\dagger }+\sum\limits_{j=1}^\infty \left( -1\right) ^j({\Bbb C}%
_n^{\dagger }{\Bbb D}_n^{\dagger })^j  \label{3-19}
\end{equation}

The spectral decomposition of the intermediate operator $\Theta _n^{\dagger
},$\ i.e. a set of right (left) generalized eigenvectors $|\tilde{u}%
_{n\alpha })$, ($(u_{n\alpha }|$) satisfying 
\begin{equation}
(\tilde{u}_{n\alpha }|u_{m\beta })=\delta _{nm}\delta _{\alpha \beta
},\qquad \Theta _n^{\dagger }=\sum\limits_\alpha z_{n\alpha }|\tilde{u}%
_{n\alpha })(u_{n\alpha }|,\qquad {\Bbb P}_n^{\dagger }=\sum\limits_\alpha |%
\tilde{u}_{n\alpha })(u_{n\alpha }|.  \label{4-19}
\end{equation}
where $\alpha $\ and $\beta $\ are discrete or continuous indexes. In the
later case the sums in the previous expression should be replaced by
integrals and the Kronecker by Dirac deltas.

The spectral decomposition of ${\Bbb L}^{\dagger }$\ can be obtained from
the spectral decomposition of the intermediate operator $\Theta ^{\dagger }.$%
\ From equations (\ref{3-20a}) and (\ref{4-19}) it follows that 
\begin{equation}
{\Bbb L}^{\dagger }=\sum_{n\alpha }z_{n\alpha }\,|\tilde{f}_{n\alpha
})\,(f_{n\alpha }|,
\end{equation}
where 
\begin{equation}
|\tilde{f}_{n\alpha })=(\Omega ^{\dagger })^{-1}\,|\tilde{u}_{n\alpha
})\qquad \,\left( f_{n\alpha }\right| =\left( u_{n\alpha }\right| \Omega
^{\dagger }.  \label{4-22}
\end{equation}

The time evolution of a state functional, governed by the generalized
Liouville-Von Newmann equation 
\begin{equation}
-i\frac d{dt}(\rho _t|=(\rho _t|\,{\Bbb L}^{\dagger },  \label{3-22a}
\end{equation}
is given by 
\begin{equation}
(\rho _t|=\sum_{n\alpha }e^{i\cdot z_{n\alpha }\cdot t}\,(\rho _0|\tilde{f}%
_{n\alpha })(f_{n\alpha }|.  \label{3-23}
\end{equation}

\subsection{Perturbative solutions and $\lambda ^2t$ approximation.}

If we replace the zero order approximation for ${\Bbb C}_n^{\dagger }$ and $%
{\Bbb D}_n^{\dagger }$ (i.e. ${\Bbb C}_n^{{\Bbb \dagger }\left( 0\right) }=%
{\Bbb D}_n^{{\Bbb \dagger }\left( 0\right) }=0$) in the right hand side of
equations (\ref{3-20}) and (\ref{3-21}), we can obtain the first order
approximations: 
\begin{eqnarray}
(\alpha |{\Bbb C}_n^{{\Bbb \dagger }\left( 1\right) }|\beta ) &=&\frac{-1}{%
\omega _\beta -\omega _\alpha \pm i0}(\alpha \mid {\Bbb P}_n^{{\Bbb \dagger }%
}{\Bbb L}_1^{{\Bbb \dagger }}{\Bbb Q}_n^{{\Bbb \dagger }}\mid \beta ),\qquad
n_\beta 
{> \atop <}
n_\alpha  \label{3-19a} \\
(\alpha |{\Bbb D}_n^{{\Bbb \dagger }\left( 1\right) }|\beta ) &=&\frac 1{%
\omega _\beta -\omega _\alpha \pm i0}(\alpha \mid {\Bbb Q}_n^{{\Bbb \dagger }%
}{\Bbb L}_1^{{\Bbb \dagger }}{\Bbb P}_n^{{\Bbb \dagger }}\mid \beta ),\qquad
n_\beta 
{< \atop >}
n_\alpha  \label{3-19b}
\end{eqnarray}

With ${\Bbb C}_n^{{\Bbb \dagger }\left( 1\right) }$ and ${\Bbb D}_n^{{\Bbb %
\dagger }\left( 1\right) }$ it is possible to obtain the intermediate
operators $\Theta _n^{\dagger }$ up to second order, using equation (\ref
{3-22}) 
\begin{equation}
\Theta _n^{\dagger \left( 2\right) }={\Bbb P}_n^{\dagger }{\Bbb L}^{\dagger }%
{\Bbb P}_n^{\dagger }+{\Bbb P}_n^{\dagger }{\Bbb C}_n^{\dagger \left(
1\right) }{\Bbb L}_1^{\dagger }{\Bbb P}_n^{\dagger }  \label{3-19c}
\end{equation}

From equations (\ref{3-20a}), (\ref{3-19}), (\ref{3-19a}) and (\ref{3-19b}),
we can compute $\Omega ^{\dagger }$\ and $(\Omega ^{\dagger })^{-1}$ up to
first order

\begin{eqnarray}
\Omega ^{\dagger (1)} &=&\sum\limits_n({\Bbb P}_n^{\dagger }+{\Bbb C}%
_n^{\dagger \left( 1\right) })  \nonumber \\
(\Omega ^{\dagger })^{-1(1)} &=&\sum\limits_n({\Bbb P}_n^{\dagger }+{\Bbb D}%
_n^{\dagger \left( 1\right) })  \label{4-24}
\end{eqnarray}

The first order expressions for ${\Bbb C}_n^{{\Bbb \dagger }}$ and ${\Bbb D}%
_n^{\dagger }$ given in equations (\ref{3-19a}) and (\ref{3-19b}) can be
replaced in the right hand side of equations (\ref{3-20}) and (\ref{3-21})
to obtain the next order approximation. In this way, through the computation
of the eigenvalues and eigenvectors of the intermediate operator $\Theta
^{\dagger }$, it is possible to obtain the eigenvalues and eigenvectors of $%
{\Bbb L}^{\dagger }$ as a power expansion in the interaction parameter 
\begin{eqnarray}
z_{n\alpha } &=&z_{n\alpha }^{(0)}+z_{n\alpha }^{(1)}+z_{n\alpha
}^{(2)}+\cdot \cdot \cdot  \nonumber \\
|\tilde{f}_{n\alpha }) &=&|\tilde{f}_{n\alpha }^{(0)})+|\tilde{f}_{n\alpha
}^{(1)})+|\tilde{f}_{n\alpha }^{(2)})+\cdot \cdot \cdot  \label{4-28} \\
(f_{n\alpha }| &=&(f_{n\alpha }^{(0)}|+(f_{n\alpha }^{(1)}|+(f_{n\alpha
}^{(2)}|+\cdot \cdot \cdot  \nonumber
\end{eqnarray}

Taking into account equations (\ref{4-22}) and (\ref{4-24}) relating the
eigenvectors of $\Theta ^{\dagger }$\ and ${\Bbb L}^{\dagger }$\ we obtain 
\begin{eqnarray}
|\tilde{f}_{n\alpha }) &=&|\widetilde{u}_{n\alpha }^{(0)})+{\Bbb D}%
_n^{\dagger (1)}|\widetilde{u}_{n\alpha }^{(0)})+|\widetilde{u}_{n\alpha
}^{(1)})+\cdot \cdot \cdot  \nonumber \\
(f_{n\alpha }| &=&(u_{n\alpha }^{(0)}|+(u_{n\alpha }^{(0)}|{\Bbb C}%
_n^{\dagger (1)}+(u_{n\alpha }^{(1)}|+\cdot \cdot \cdot  \label{4-29}
\end{eqnarray}

Replacing (\ref{4-28}) and (\ref{4-29}) in (\ref{3-23}) we obtain the
following expression for the time evolution 
\begin{eqnarray*}
(\rho _t| &=&\sum\limits_{n\alpha }\exp \left[ i(z_{n\alpha
}^{(0)}+z_{n\alpha }^{(1)}+z_{n\alpha }^{(2)}+\cdot \cdot \cdot )t\right]
\times \\
&&\times (\rho _0|\left[ |\widetilde{u}_{n\alpha }^{(0)})(u_{n\alpha
}^{(0)}|+|\widetilde{u}_{n\alpha }^{(0)})(u_{n\alpha }^{(0)}|{\Bbb C}%
_n^{\dagger (1)}+{\Bbb D}_n^{\dagger (1)}|\widetilde{u}_{n\alpha
}^{(0)})(u_{n\alpha }^{(0)}|+|\widetilde{u}_{n\alpha }^{(1)})(u_{n\alpha
}^{(0)}|+|\widetilde{u}_{n\alpha }^{(0)})(u_{n\alpha }^{(1)}|+\cdot \cdot
\cdot \right] .
\end{eqnarray*}

If we omit first order contributions coming from the eigenvectors and third
order contributions from the eigenvalues, the previous expression has the
following approximated form 
\begin{equation}
(\rho _t|\cong \sum\limits_{n\alpha }\exp \left[ i(z_{n\alpha
}^{(0)}+z_{n\alpha }^{(1)}+z_{n\alpha }^{(2)})t\right] \,(\rho _0|\widetilde{%
u}_{n\alpha }^{(0)})(u_{n\alpha }^{(0)}|.  \label{4-30}
\end{equation}

As we omitted first order terms in the eigenvectors, a necessary condition
for equation (\ref{4-30}) to be valid is 
\begin{equation}
\lambda <<1,  \label{4-31}
\end{equation}
where $\lambda $\ is the interaction parameter. Moreover, as we considered
the eigenvalues up to second order, a second condition involving the
possible values of time is necessary for equation (\ref{4-30}) to be valid 
\begin{equation}
\lambda ^3t<<1,  \label{4-32}
\end{equation}
which together with (\ref{4-31}) gives 
\begin{equation}
t%
{< \atop \sim }
\lambda ^{-2}.  \label{4-33}
\end{equation}

In summary, the time evolution is given by the equation (\ref{4-30}) if the
interaction is small ($\lambda <<1$) and the time is not too large ($t%
{< \atop \sim }
\lambda ^{-2}$).

Conditions (\ref{4-31}) and (\ref{4-33}) appear in the literature as the ''$%
\lambda ^2t$\ approximation''.

\section{Friedrichs Model}

\subsection{Observables, states and degrees of correlation}

The Hamiltonian of Friedrichs model is 
\begin{eqnarray}
H &=&H_0+V,  \nonumber \\
H_0 &=&m|1\rangle \langle 1|+\int\limits_0^\infty d\omega \,\omega \,|\omega
\rangle \langle \omega |, \\
V &=&\int\limits_0^\infty d\omega \,V_\omega \,(|\omega \rangle \langle
1|+|1\rangle \langle \omega |).  \nonumber
\end{eqnarray}

Where the vectors $|1\rangle $ and $|\omega \rangle $ form a complete
orthonormal set 
\begin{equation}
\langle 1|1\rangle =1,\qquad \langle 1|\omega \rangle =\langle \omega
|1\rangle =0,\qquad \langle \omega |\omega ^{\prime }\rangle =\delta \left(
\omega -\omega ^{\prime }\right) ,\qquad I=|1\rangle \langle
1|+\int\limits_0^\infty d\omega \,|\omega \rangle \langle \omega |.
\end{equation}

Let us consider the following definitions 
\begin{equation}
|1)\equiv |1\rangle \langle 1|,\quad |\omega )\equiv |\omega \rangle \langle
\omega |,\quad |1\,\omega )\equiv |1\rangle \langle \omega |,\quad |\omega
\,1)\equiv |\omega \rangle \langle 1|,\quad |\omega \,\omega ^{\prime
})\equiv |\omega \rangle \langle \omega ^{\prime }|.  \label{4-3}
\end{equation}

Any element $O$ belonging to the space of observables ${\cal O}$, can be
written in terms of the operators defined in (\ref{4-3})

\begin{equation}
|O)=O_1|1)+\int d\omega \,O_\omega |\omega )+\int d\omega \,O_{1\omega
}|1\,\omega )+\int d\omega \,O_{\omega 1}|\omega \,1)+\int d\omega \int
d\omega ^{\prime }\,O_{\omega \omega ^{\prime }}|\omega \,\omega ^{\prime }).
\end{equation}

Notice that we explicitly included a diagonal singularity through the term $%
\int d\omega \,O_\omega |\omega )$.

We also consider the states as functionals acting on the operators. For this
purpose it is convenient to define a set of functionals $(1|$, $(\omega |$, $%
(1\omega |$, $(\omega 1|$ and $(\omega \omega ^{\prime }|$ with the
following properties 
\begin{equation}
(1|O)=O_1,\quad (\omega |O)=O_\omega ,\quad (1\omega |O)=O_{1\omega },\quad
(\omega 1|O)=O_{\omega 1},\quad (\omega \omega ^{^{\prime }}|O)=O_{\omega
\omega ^{^{\prime }}}.  \label{4-5}
\end{equation}
or equivalently 
\begin{eqnarray}
(1|1) &=&1,\quad (1|\omega )=(1|\omega \,1)=(1|1\,\omega )=(1|\omega
\,\omega ^{\prime })=0,  \nonumber \\
(\omega |\omega ^{\prime }) &=&\delta (\omega -\omega ^{\prime }),\quad
(\omega |1)=(\omega |1\,\omega ^{\prime })=(\omega |\omega ^{\prime
}\,1)=(\omega |\omega ^{\prime }\,\omega ^{\prime \prime })=0,  \nonumber \\
(1\,\omega |1\,\omega ^{\prime }) &=&\delta (\omega -\omega ^{\prime
}),\quad (1\,\omega |1)=(1\,\omega |\omega ^{\prime })=(1\,\omega |\omega
^{\prime }\,1)=(1\,\omega |\omega ^{\prime }\,\omega ^{\prime \prime })=0, \\
(\omega \,1|\omega ^{\prime }\,1) &=&\delta (\omega -\omega ^{\prime
}),\quad (\omega \,1|1)=(\omega \,1|\omega ^{\prime })=(\omega \,1|1\,\omega
^{\prime })=(\omega \,1|\omega ^{\prime }\,\omega ^{\prime \prime })=0, 
\nonumber \\
(\omega \omega ^{\prime }|\alpha \alpha ^{\prime }) &=&\delta (\omega
-\alpha )\delta (\omega ^{\prime }-\alpha ^{\prime }),\quad (\omega \omega
^{\prime }|1)=(\omega \omega ^{\prime }|1\alpha )=(\omega \omega ^{\prime
}|\alpha 1)=(\omega \omega ^{\prime }|\alpha )=0.  \nonumber
\end{eqnarray}

In terms of these functionals, we assume that any element $(\rho |$ of the
space of states ${\cal S\subset O}^{\times }$ can be written as 
\begin{equation}
(\rho |=\rho _1^{*}\,(1|+\int d\omega \,\rho _\omega ^{*}\,(\omega |+\int
d\omega \,\rho _{1\omega }^{*}\,(1\,\omega |+\int d\omega \,\rho _{\omega
1}^{*}\,(\omega \,1|+\int d\omega \int d\omega ^{\prime }\,\rho _{\omega
\omega ^{\prime }}^{*}\,(\omega \,\omega ^{\prime }|
\end{equation}

where 
\begin{equation}
\rho _1^{*}=\rho _1,\qquad \rho _\omega ^{*}=\rho _\omega ,\qquad \rho
_{1\omega }^{*}=\rho _{\omega 1},\qquad \rho _{\omega \omega ^{\prime
}}^{*}=\rho _{\omega ^{\prime }\omega },  \label{4-8}
\end{equation}

\begin{equation}
\rho _1^{*}+\int d\omega \,\rho _\omega ^{*}=1  \label{4-9}
\end{equation}

Equations (\ref{4-8}) are the conditions for $\rho $ to be a positive
functional, while (\ref{4-9}) is a consequence of the total probability
condition $(\rho |I)=1$. In what follows $(\rho |I)$ will be called the {\it %
generalized trace}{\bf \ }of the state $\rho $ ($|I)\equiv |1)+\int d\omega
|\omega )$ is the {\it identity operator} in ${\cal O}$).

By using the basis for states and observables, defined through (\ref{4-3})
and (\ref{4-5}), we can also write 
\begin{eqnarray}
{\Bbb L}_0^{\dagger } &=&\int d\omega \,(m-\omega )|1\,\omega )(1\,\omega
|+\int d\omega \,(\omega -m)|\omega \,1)(\omega \,1|+\int d\omega \int
d\omega ^{\prime }\,(\omega -\omega ^{\prime })|\omega \,\omega ^{\prime
})(\omega \,\omega ^{\prime }|, \\
{\Bbb L}_1^{\dagger } &=&\int d\omega V_\omega [|\omega 1)-|1\omega
)](1|+\int d\omega V_\omega [|1\omega )-|\omega 1)](\omega |+\int d\omega
[-V_\omega |1)+\int d\omega ^{\prime }V_{\omega ^{\prime }}|\omega ^{\prime
}\omega )](1\omega |+  \nonumber \\
&&+\int d\omega [V_\omega |1)-\int d\omega ^{\prime }V_{\omega ^{\prime
}}|\omega \omega ^{\prime })](\omega 1|+\int d\omega \int d\omega ^{\prime
}[V_\omega |1\omega ^{\prime })-V_{\omega ^{\prime }}|\omega 1)](\omega
\omega ^{\prime }|,  \nonumber
\end{eqnarray}
where ${\Bbb L}_0^{\dagger }$ and ${\Bbb L}_1^{\dagger }$ are the ''free''
and ''interaction'' parts of the Liouville-Von Newmann operator acting on $%
{\cal O},$ i.e. 
\begin{equation}
{\Bbb L}_0^{\dagger }O\equiv \left[ H_0,O\right] \;\;\;\;\;{\Bbb L}%
_1^{\dagger }O\equiv \left[ V,O\right] \;\;\;\;\;O\in {\cal O}
\end{equation}

The diagonal and off-diagonal projectors, acting on ${\cal O}$, are defined
by 
\begin{eqnarray}
{\Bbb P}_0^{\dagger }\equiv |1)(1|+\int d\omega |\omega )(\omega |, 
\nonumber \\
{\Bbb Q}_0^{\dagger }\equiv \int d\omega |1\,\omega )(1\,\omega |+\int
d\omega |\omega \,1)(\omega \,1|+\int d\omega \,d\omega ^{\prime }|\omega
\,\omega ^{\prime })(\omega \,\omega ^{\prime }|
\end{eqnarray}

The off-diagonal projector ${\Bbb Q}_0^{\dagger }$ can be decomposed into 
\begin{equation}
{\Bbb Q}_0^{\dagger }={\Bbb P}_1^{\dagger }+{\Bbb P}_2^{\dagger },\quad 
{\Bbb P}_1^{\dagger }\equiv \int d\omega |1\,\omega )(1\,\omega |+\int
d\omega |\omega \,1)(\omega \,1|,\quad {\Bbb P}_2^{\dagger }\equiv \int
d\omega \int d\omega ^{\prime }|\omega \,\omega ^{\prime })(\omega \,\omega
^{\prime }|.
\end{equation}

${\Bbb P}_0^{\dagger }$, ${\Bbb P}_1^{\dagger }$ and ${\Bbb P}_2^{\dagger }$
are the projectors corresponding to degrees of correlation zero, one and two
respectively, i.e. 
\begin{eqnarray}
{\Bbb P}_0^{\dagger }\,({\Bbb L}_1^{\dagger })^0{\Bbb \,P}_0^{\dagger } &=&%
{\Bbb P}_0^{\dagger 2}\neq 0.  \nonumber \\
{\Bbb P}_0^{\dagger }\,({\Bbb L}_1^{\dagger })^0{\Bbb \,P}_1^{\dagger } &=&%
{\Bbb P}_0^{\dagger }\,{\Bbb P}_1^{\dagger }=0,\qquad {\Bbb P}_0^{\dagger }\,%
{\Bbb L}_1^{\dagger }\,{\Bbb P}_1^{\dagger }\neq 0.  \nonumber \\
{\Bbb P}_0^{\dagger }\,({\Bbb L}_1^{\dagger })^0\,{\Bbb P}_2^{\dagger } &=&%
{\Bbb P}_0^{\dagger }\,{\Bbb P}_2^{\dagger }=0,\qquad {\Bbb P}_0^{\dagger }\,%
{\Bbb L}_1^{\dagger }\,{\Bbb P}_2^{\dagger }=0,\qquad {\Bbb P}_0^{\dagger
}\,({\Bbb L}_1^{\dagger })^2\,{\Bbb P}_2^{\dagger }\neq 0.
\end{eqnarray}

\subsection{Creation, Destruction and Intermediate Operators}

Using the equation (\ref{3-19a}) and (\ref{3-19b}) for the Friedrichs model
we obtain for the creation and destruction operator up to second order are:

\begin{eqnarray}
{\Bbb C}_0^{\dagger (1)} &=&\int d\omega \,V_\omega \left[ \frac{%
|1)(1\,\omega |}{m-\omega +i0}-\frac{|1)(\omega \,1|}{\omega -m+i0}\right] ,
\nonumber \\
{\Bbb C}_1^{\dagger (1)} &=&\int d\omega V_\omega \left[ \frac{|\omega
1)[(1|-(\omega |]}{\omega -m+i0}-\frac{|1\omega )[(1|-(\omega |]}{m-\omega
+i0}\right] +\int d\omega d\omega ^{\prime }\left[ \frac{V_{\omega ^{\prime
}}|\omega 1)(\omega \omega ^{\prime }|}{m-\omega ^{\prime }+i0}-\frac{%
V_\omega |1\omega ^{\prime })(\omega \omega ^{\prime }|}{\omega -m+i0}%
\right] ,  \nonumber \\
{\Bbb C}_2^{\dagger (1)} &=&\int d\omega \,d\omega ^{\prime }\left[ \frac{%
V_{\omega ^{\prime }}}{\omega ^{\prime }-m+i0}|\omega ^{\prime }\,\omega
)(1\,\omega |+\frac{V_{\omega ^{\prime }}}{\omega ^{\prime }-m-i0}|\omega
\,\omega ^{\prime })(\omega \,1|\right] ,  \label{4-16} \\
{\Bbb D}_0^{\dagger (1)} &=&\int d\omega \,V_\omega \left[ \frac{|1\,\omega
)(1|}{m-\omega +i0}+\frac{|\omega \,1)(1|}{m-\omega -i0}\right] -\int
d\omega \,V_\omega \left[ \frac{|1\,\omega )(\omega |}{m-\omega +i0}+\frac{%
|\omega \,1)(\omega |}{m-\omega -i0}\right] ,  \nonumber \\
{\Bbb D}_1^{\dagger (1)} &=&\int d\omega \,V_\omega \left[ \frac{|1)(\omega
\,1|}{\omega -m+i0}-\frac{|1)(1\,\omega |}{m-\omega +i0}\right] +\int
d\omega \,d\omega ^{\prime }\,V_{\omega ^{\prime }}\left[ \frac{|\omega
\,\omega ^{\prime })(\omega \,1|}{m-\omega ^{\prime }+i0}-\frac{|\omega
^{\prime }\,\omega )(1\,\omega |}{\omega ^{\prime }-m+i0}\right] ,  \nonumber
\\
{\Bbb D}_2^{\dagger (1)} &=&\int d\omega \,d\omega ^{\prime }\left[ \frac{%
V_\omega |1\,\omega ^{\prime })(\omega \,\omega ^{\prime }|}{\omega -m+i0}-%
\frac{V_{\omega ^{\prime }}|\omega \,1)(\omega \,\omega ^{\prime }|}{%
m-\omega +i0}\right] .  \nonumber
\end{eqnarray}

Then, the intermediate operator $\Theta _n^{\dagger }$, up to second order,
can be obtained using equation (\ref{3-19c})

\begin{eqnarray}
\Theta _0^{\dagger \left( 2\right) } &=&2\pi i\,V_m^2\left| 1\right) \left[
\left( 1\right| -\left( m\right| \right] ,\;\;\;\;\;\left( m\right| \equiv
\left( \omega \right| _{\omega =m}  \nonumber \\
\Theta _1^{\dagger }{}^{(2)} &=&\int d\omega [m-\omega -\beta ]|1\,\omega
)(1\,\omega |+\int d\omega [\omega -m-\beta ^{*}]|\omega \,1)(\omega \,1|+ 
\nonumber \\
&&+\int d\omega \,d\omega ^{^{\prime }}\,V_\omega \,V_{\omega ^{\prime
}}\,\left[ \frac{|1\,\omega ^{\prime })(1\,\omega |}{m-\omega +i0}+\frac{%
|\omega ^{\prime }\,1)(\omega \,1|}{m-\omega +i0}+\frac{|\omega ^{\prime
}\,1)(1\,\omega |}{m-\omega ^{\prime }-i0}\right] +  \nonumber \\
&&+\int d\omega \,d\omega ^{\prime }\,V_\omega \,V_{\omega ^{\prime }}\left[ 
\frac{|\omega ^{\prime }\,1)(\omega \,1|}{\omega ^{\prime }-m+i0}+\frac{%
|1\,\omega ^{\prime })(\omega \,1|}{\omega -m+i0}+\frac{|1\,\omega ^{\prime
})(\omega \,1|}{\omega ^{\prime }-m-i0}\right] ,  \nonumber \\
\beta &\equiv &\int_0^\infty \frac{d\omega \,V_\omega ^2}{\omega -m+i0}%
,\quad Re\beta =\int_0^\infty d\omega \,V_\omega ^2\,P\left( \frac 1{\omega
-m}\right) ,\quad Im\beta =-\pi \cdot V_m^2  \nonumber \\
\Theta _2^{\dagger }{}^{(2)} &=&\int d\omega d\omega ^{\prime }(\omega
^{^{\prime }}-\omega )|\omega \,\omega ^{\prime })(\omega \,\omega ^{\prime
}|+  \nonumber \\
&&+\int d\omega \,d\omega ^{^{\prime }}\,d\omega ^{\prime \prime }\,\left[ 
\frac{V_\omega \,V_{\omega ^{\prime \prime }}\,|\omega ^{\prime \prime
}\,\omega ^{\prime })(\omega \,\omega ^{\prime }|}{\omega ^{\prime \prime
}-m+i0}-\frac{V_{\omega ^{\prime }}\,V_{\omega ^{\prime \prime }}|\omega
\,\omega ^{\prime \prime })(\omega \,\omega ^{\prime }|}{\omega ^{\prime
\prime }-m-i0}\right] .  \label{4-18}
\end{eqnarray}

\subsection{Generalized spectral decomposition and time evolution.}

From the explicit form of $\Theta _n^{\dagger }$ given up to second order in
equations (\ref{4-18}) for Friedrichs model, the generalized eigenvectors
and eigenvalues can be computed up to zero and second order respectively.
The results are shown in the following table

\begin{equation}
\begin{tabular}{|c|c|c|c|}
\hline
$\Theta _n^{\dagger }$ & $z_{n\alpha }$ & $|\tilde{u}_{n\alpha })$ & $%
(u_{n\alpha }|$ \\ \hline
$\Theta _0^{\dagger }$ & $
\begin{array}{l}
z_\omega =0 \\ 
z_1=2\pi i\,V_m^2
\end{array}
$ & $
\begin{array}{l}
|\tilde{u}_\omega )=\delta (\omega -m)|1)+|\omega ) \\ 
|\tilde{u}_1)=|1)
\end{array}
$ & $
\begin{array}{l}
(u_\omega |=(\omega | \\ 
(u_1|=(1|-(\omega =m|
\end{array}
$ \\ \hline
$\Theta _1^{\dagger }$ & $
\begin{array}{l}
z_{1\omega }=m-\omega -\beta \\ 
z_{\omega 1}=\omega -m-\beta ^{*}
\end{array}
$ & $
\begin{array}{l}
|\tilde{u}_{1\omega })=|1\omega ) \\ 
|\tilde{u}_{\omega 1})=|\omega 1)
\end{array}
$ & $
\begin{array}{l}
(u_{1\omega }|=(1\omega | \\ 
(u_{\omega 1}|=(\omega 1|
\end{array}
$ \\ \hline
$\Theta _2^{\dagger }$ & $z_{\omega \omega ^{\prime }}=\omega -\omega
^{\prime }$ & $|\tilde{u}_{\omega \omega ^{\prime }})=|\omega \omega
^{\prime })$ & $(u_{\omega \omega ^{^{\prime }}}|=(\omega \omega ^{\prime }|$
\\ \hline
\end{tabular}
\label{4-20}
\end{equation}

The generalized eigenvalues and eigenvectors given in the previous
expression can be replaced in the equation (\ref{4-30}), to obtain the
following time evolution 
\begin{eqnarray*}
(\rho _t|1) &\cong &e^{-2\pi V_m^2\,t}\,(\rho _0|1), \\
(\rho _t|\omega ) &\cong &(\rho _0|\omega )+\left[ 1-e^{-2\pi
V_m^2\,t}\right] \,(\rho _0|1)\,\delta (\omega -m), \\
(\rho _t|1\omega ) &\cong &e^{i(m-\omega -\beta )\,t}(\rho _0|1\omega ), \\
(\rho _t|\omega 1) &\cong &e^{i(\omega -m-\beta ^{*})\,t}(\rho _0|\omega 1),
\\
(\rho _t|\omega \omega ^{\prime }) &\cong &e^{i(\omega -\omega ^{\prime
})t}(\rho _0|\omega \omega ^{\prime }).
\end{eqnarray*}

The first equation shows the decay of the discrete component $(\rho _t|1)$
of the state, with a rate $2\pi V_m^2$. Simultaneously, there is a growing
term in the continuous distribution $(\rho _t|\omega )$, with a sharp peak
for the energy $\omega =m$ of the decaying mode.

\section{Conclusions}

We extended the formalism of subdynamics to the functional approach of
quantum mechanics, in which the states are represented by functionals acting
on the operators representing observables.

The generalized spectral decomposition is obtained through an intermediate
superoperator $\Theta ^{\dagger }$ isospectral to the Liouville-Von Newmann
superoperator ${\Bbb L}^{\dagger }$ (${\Bbb L}^{\dagger }=\Omega ^{\dagger
}\Theta ^{\dagger }\Omega ^{-1\dagger }$). The small denominators appearing
in the perturbative expansions due to the continuous spectrum are
regularized by the ''$i\varepsilon $-rule'' (a time ordering prescription in
which increasing (decreasing) of correlations is future (past) oriented).
Due to this time ordering rule, $\Theta ^{\dagger }$ and therefore ${\Bbb L}%
^{\dagger }$ may have complex eigenvalues.

Considering eigenvalues up to second order and eigenvectors up to zero
order, the time evolution is given by 
\[
(\rho _t|\cong \sum\limits_{n\alpha }\exp \left[ i(z_{n\alpha
}^{(0)}+z_{n\alpha }^{(1)}+z_{n\alpha }^{(2)})t\right] \,(\rho _0|\widetilde{%
u}_{n\alpha }^{(0)})(u_{n\alpha }^{(0)}|, 
\]
where $|\widetilde{u}_{n\alpha }^{(0)})$ and $(u_{n\alpha }^{(0)}|$ are
generalized right and left eigenvector of $\Theta _n^{\dagger }$ computed up
to zero order. For the previous expression to be valid, it is necessary that
the interaction parameter be small and the time not too large, i.e. $\lambda
\ll 1$ and $t%
{< \atop \sim }
\lambda ^{-2}$.

When this procedure is applied to the Friedrichs model, we obtain 
\begin{eqnarray*}
(\rho _t|1) &\cong &e^{-2\pi V_m^2\,t}\,(\rho _0|1), \\
(\rho _t|\omega ) &\cong &(\rho _0|\omega )+\left[ 1-e^{-2\pi
V_m^2\,t}\right] \,(\rho _0|1)\,\delta (\omega -m), \\
(\rho _t|1\omega ) &\cong &e^{i(m-\omega -\beta )\,t}(\rho _0|1\omega ), \\
(\rho _t|\omega 1) &\cong &e^{i(\omega -m-\beta ^{*})\,t}(\rho _0|\omega 1),
\\
(\rho _t|\omega \omega ^{\prime }) &\cong &e^{i(\omega -\omega ^{\prime
})t}(\rho _0|\omega \omega ^{\prime }).
\end{eqnarray*}

The first equation shows the decay of the discrete component $(\rho _t|1)$
of the state, with a rate $2\pi V_m^2$. Simultaneously, there is a growing
term in the continuous distribution $(\rho _t|\omega )$, with a sharp peak
for the energy $\omega =m$ of the decaying mode.

It is interesting to note that both decaying and growing terms are purely
exponential. This may appear at first sight as a contradiction with the well
known Zeno and Khalfin effects, which are deviations from exponential decays
for small and big times. However, the previous expressions are not valid
approximations for very big times. Moreover, Zeno effect implies $\frac d{dt}%
(\rho _t|1)_{t=0}$. In our approximation, if we compute this derivative we
obtain that it is of second order in the interaction parameter. As we
neglected this order in the approximation, this result is not in
contradiction with Zeno effect.

In spite of the fact that the complex spectral decomposition can be obtained
analytically for the Friedrichs model \cite{22}\cite{23}\cite{24}, this
paper shows that this approach is potentially suitable to deal with more
complicated decaying processes,where it is impossible to obtain exact
solutions

\begin{center}
{\bf ACKNOWLEDGMENTS.}
\end{center}

This work was partially supported by Grant No. CI1-CT94-0004 of the European
Community, Grant No. PID-0150 of CONICET (National Research Council of
Argentina), Grant No. EX-198 of Buenos Aires University, Grant No. 12217/1
of Fundaci\'{o}n Antorchas, and also a Grant from Foundation pour la
Recherche Foundamentale OLAM.

\end{document}